\documentclass[12pt,letterpaper]{article}          
\usepackage{osajnl}
\usepackage[draft]{hyperref} 

\def \apj {ApJ}
\def \apjs {ApJS}

\begin{document}

\title{Polarimetric Calibration of Large-Aperture Telescopes II: The
  sub-aperture method} 

\author{Hector Socas-Navarro}

\address{3450 Mitchell Ln, Boulder CO, 80307-3000}

\email{navarro@ucar.edu}

\begin{abstract}
A new method for absolute polarimetric calibration of large telescopes is
presented. The proposed method is highly accurate and is based on the
calibration of a small sub-aperture, which is then extended
to the full system by means of actual observations of an astronomical
source. The calibration procedure is described in detail along with numerical
simulations that explore its robustness and accuracy. The advantages and
disadvantages of this technique with respect to other possible alternatives
are discussed.
\end{abstract}


\maketitle 

\section{Introduction}
\label{sec:intro}

Polarimetry has become a fundamental part of solar physics and is rapidly
gaining importance in many other fields of astrophysics\cite{WS02}. One of
the most challenging problems that one faces when doing astronomical
polarimetry is the determination of the instrumental polarization introduced
by the telescope in the observed beam. There are two approaches that have
been used to deal with this problem. The first one is to place calibration
optics at the telescope entrance, which allows one to feed light in a
known state of polarization into the system. The measured Stokes parameters
can then be used to determine the telescope response with a very high degree
of accuracy. Two solar telescopes currently use this strategy: The Dunn Solar
Telescope (DST), at the Sacramento Peak observatory (NM, USA) operated by the
National Solar Observatories, and the German Vacuum Tower Telescope (VTT) at
the Observatorio del Teide of the Instituto de Astrof\' \i sica de Canarias
(Spain). The Swedish Solar Tower at
the Observatorio del Roque de Los Muchachos (also of the Instituto de 
Astrof\' \i sica de Canarias) will soon have this capability, as well.

The second approach to determine the telescope polarimetric response is to
observe relatively well-known astronomical sources of polarization. This is
sometimes done in night-time astronomy by observing ``standard stars''. In
solar physics, this may be accomplished with observations of a sunspot.
The procedure\cite{S04} considers Milne-Eddington 
fits (MEF) to sunspot umbra spectra, from which the elements of the telescope
Mueller matrix are adjusted to minimize the residuals of the MEF to the
observed data.

It is generally desirable to have a purely instrumental telescope
calibration, so that no {\it a priori} assumptions
on the observed targets are needed. Unfortunately, the construction of
calibration optics for a large aperture (larger than
$\sim$1~m) is impractical. This had not been an issue in the past because
existing solar 
telescopes have apertures that do not exceed 1~m in diameter, whereas
larger night-time telescopes did not require absolute polarimetric accuracies
higher than what can be routinely achieved by means of standard stars. This
situation, however, is rapidly changing due to the ever increasing accuracy
demands of astronomical polarimetry and the development of new large-aperture
solar telescopes, like the German telescope GREGOR\cite{VvdLK+03} or the
Advanced Technology Solar Telescope\cite{KRH+02} \cite{KRK+03} (ATST).

A previous paper\cite{SN04c} 
introduced the ``beam-expansion'' method (BEM) for the instrumental
calibration of a large telescope. This work presents a 
different procedure, the sub-aperture method, which has
some interesting 
advantages over the BEM (at least for the particular case of the ATST
design). We shall hereafter consider the problem of calibrating the ATST, but
the proposed procedure may be easily applicable to other telescopes as well.

The sub-aperture method is based on the following basic idea. One starts by
effectively turning the ATST into a small-aperture telescope. This is done by
inserting a mask into the beam at any pupil
image. The telescope sub-aperture is calibrated as usual by means of
calibration polarization optics above the primary mirror. An astronomical
polarization source (e.g., a sunspot) is observed through the sub-aperture
and then again with the full aperture. The two observations are only a few
seconds apart in time. The sub-aperture calibration is used to remove
instrumental polarization in the first observation, which allows one to
determine the actual 
Stokes vectors emerging from the astronomical source. Once these input
vectors are known, the full aperture observation can be used to calibrate the
telescope. 

\section{Requirements}
\label{sec:require}

The sub-aperture method proposed in this work requires the following
elements. Note that, for the case of the ATST, requirement~1 is probably the
most demanding one. 

\begin{enumerate}
\item A mechanical mount is needed to slide calibration optics in
  and out of the beam before the primary mirror. The calibration optics
  may have a small diameter (see next point) and consist of an achromatic
  linear polarizer and a retarder. The polarizer and retarder can rotate and
  be inserted and removed independently of one another. It is important to
  realize that the calibration optics does not need to be placed over the
  center of the aperture. It could be at any location over the primary
  mirror, e.g. near the outer edge of the
  telescope aperture. This consideration probably simplifies the design of
  the 
  mechanical mount, since it does not need to extend all the way into the
  center of the aperture. Moreover, there is an additional advantage
  for off-axis telescopes like the ATST. By having the sub-aperture closer to
  the axis of the system, rather than the aperture center, 
  the instrumental polarization of the sub-aperture is minimized. This
  is of relevance because, as discussed in \S\ref{sec:simul}, the level
  of accuracy achieved is limited by the accuracy of the sub-aperture
  calibration. 
\item The diameter of the calibration optics ($d_0$) may be as small as
  desired, 
  provided only that it fills an aperture large enough to observe the source
  of polarization. Typically this would be a sunspot. However, there may
  be times of low solar activity in which no sunspots are available. In such
  cases plage regions may be employed as sources. A suitable value for
  $d_0$ is probably between 20~cm and 40~cm (see \S\ref{sec:simul}).
\item A blocking mask that can be inserted in the beam independently of the
  calibration optics is required. The location of the mask can be any
  position in 
  the system where an image of the telescope pupil is formed. It may be
  at the entrance (e.g., the dome shutter) or, perhaps more convenient, at
  any location downstream in a collimated beam (which would make it a small
  element). The purpose of the mask is to block the rays in the light beam
  that did not pass through the calibration optics. In other words, when the
  mask is in the beam the telescope aperture is reduced to $d_0$. The
  alignment between the mask and the calibration optics does not need to be
  very accurate. We simply require that only light that passed through the
  calibration optics can get through the mask.
\end{enumerate}

\section{Calibration procedure}
\label{sec:proc}

The procedure proposed to calibrate the optical train is as follows:

\begin{enumerate}
\item Slide calibration optics into the beam
\item Slide mask into the beam. Only polarized light passes through.
\item Calibration Stokes vectors are measured for various configurations of
  the calibration optics, in the same way as it is presently done with other
  solar telescopes. If ${\bf (xt)}$ represents the Mueller matrix of the
  small-aperture telescope+instrument system, we have that:
\begin{equation}
\vec S_{cal}^{out} = {\bf (xt)} \vec S_{cal}^{in} \, .
\end{equation}
\item From (known) inputs $\vec S_{cal}^{in}$ and (measured) outputs $\vec
  S_{cal}^{out}$, determine ${\bf (xt)}$. In general the measured Mueller
  matrix, that we denote by ${\bf (xt)'}$, is affected of small errors.
\begin{equation}
\label{appr1}
  {\bf (xt)'} \simeq {\bf (xt)} \, .
\end{equation}
\item Remove calibration optics, {\it but leave mask in}.
\item Observe solar source of polarization (e.g., a sunspot) through
  mini-aperture of diameter $d_0$:
\begin{equation}
\vec S_{solar,1}^{out} = {\bf (xt)} \vec S_{solar,1}^{in} \, .
\end{equation}
\item Remove mask and repeat same observation using full aperture. Only a few
  seconds pass between the observations with and without the mask. If ${\bf
  (XT)}$ is the Mueller matrix of the full-aperture telescope+instrument
  system, we have:
\begin{equation}
\vec S_{solar,2}^{out} = {\bf (XT)} \vec S_{solar,2}^{in} \, .
\end{equation}
\item Calibrate small-aperture solar observations $\vec
  S_{solar,1}^{out}$ to obtain $\vec
  S_{solar,1}^{in}$: 
\begin{equation}
\label{appr2}
\vec S_{solar,1}^{in} \simeq \vec {S'}_{solar,1}^{in} =
{\bf (xt)'}^{-1} \vec S_{solar,1}^{out} \, .
\end{equation}
\item Use known $\vec S_{solar,1}^{in}$ and measured $\vec
  S_{solar,2}^{out}$ to determine ${\bf (XT)}$. The following assumption is
  implicit: 
\begin{equation}
\label{approx12}
\vec S_{solar,1}^{in} \simeq \vec S_{solar,2}^{in} \, .
\end{equation}
\end{enumerate}

There are two important assumptions in the procedure as outlined
above. The first one is given by Eq~(\ref{appr1}) and requires that the
calibration of the sub-aperture must be done with sufficient
accuracy. Possible errors in this step will affect the full aperture
calibration through Eq~(\ref{appr2}). 

The second assumption is that of Eq~(\ref{approx12}). In
order for it to be accurate, it is important that the two observations
$\vec S_{solar,1}$ and $\vec S_{solar,2}$ be taken as close in time
as possible. It is also important that the field of view
remain the same, especially when doing slit spectroscopy. Possible
discrepancies in the telescope pointing between the two observations need to
be much smaller than the spatial structure of the 
polarization source. This would probably not
be an issue when an adaptive optics system is
employed to compensate for seeing-induced wave-front distortion. However,
it is not yet clear that the adaptive optics system of the ATST will be able
to operate while in calibration mode.

The validity of these two assumptions, and the degree to which they limit the
accuracy of the telescope calibration, is explored by means of numerical
simulations in \S\ref{sec:simul}.

\section{Numerical simulations}
\label{sec:simul}

\subsection{The hound and hares tests}
\label{sec:tests}

This section describes some numerical experiments aimed at estimating the
accuracy of the sub-aperture method and its 
limitations. The simulations consist of several ``hound and hares'' type of
tests, in which one seeks to determine a telescope matrix which is known
beforehand. I have used calibrated data from the ASP instrument to simulate a
set of solar Stokes vectors ($\vec S_{solar,1}^{in}$ and $\vec
S_{solar,2}^{in}$). The simulated solar data is observed through a small
sub-aperture system (characterized by its Mueller matrix ${\bf (xt)}$) and
through the full system (with Mueller matrix ${\bf (XT)}$). 

Obviously, the actual Mueller matrix of the ATST is still unknown. Using
that of an existing telescope is probably not very helpful, since their
Mueller matrices are likely to be very different. The approach taken in this
work is to perform montecarlo simulations with a large number of randomly
chosen Mueller matrices that represent the telescope. According to reference
\cite{SN04c}, off-diagonal elements of up to $\simeq$5\% are to be expected 
due to the off-axis geometry of the ATST. Therefore, it seems appropriate to
consider matrices of the form:
\begin{eqnarray}
\label{random}
{\bf xt} = {\bf 1} + 0.05 \times {\bf \sigma}_1 \nonumber \\
{\bf XT} = {\bf 1} + 0.05 \times {\bf \sigma}_2 \, ,
\end{eqnarray}
where ${\bf 1}$ is the 4$\times$4 identity matrix and ${\bf \sigma}_i$ is a
particular realization of a 4$\times$4 matrix whose elements are random
real numbers obeying a normal distribution of width 1.

\begin{figure*}
\includegraphics[width=1.0\textwidth]{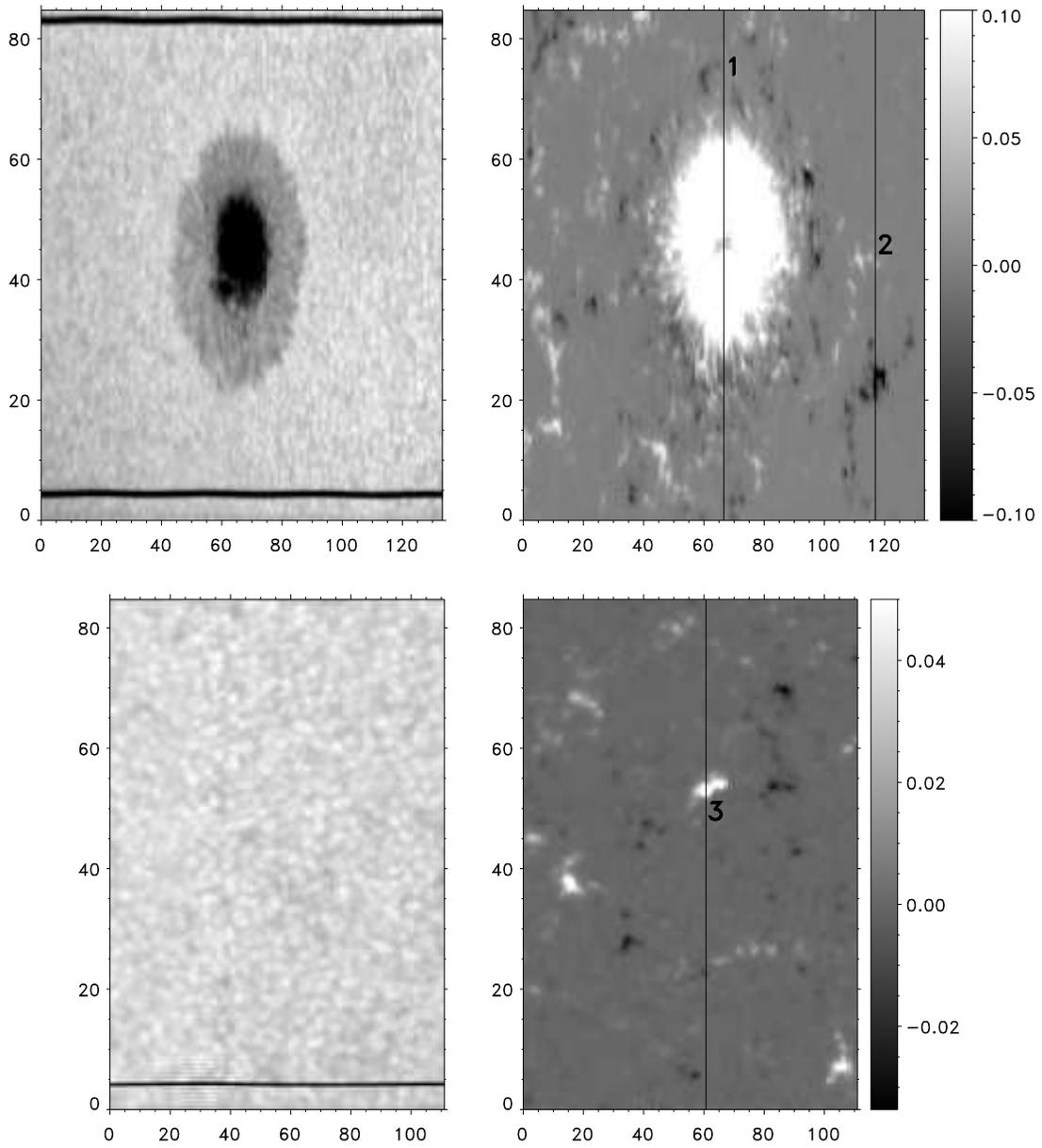}
\caption{
\label{fig:maps}
Left: Continuum intensity maps. Right: Magnetograms (degree of circular
polarization). Upper 
      panels: Map~A. Lower panels: Map~B. Scales on horizontal and vertical
      axes are in arc-seconds. The vertical lines mark the three slit
      positions used in the simulations as representative of sunspot, plage
      and network, respectively.
}
\end{figure*}

Two different ASP maps are used in the simulations. The first one
(Map~A)\cite{LTB+98} contains a sunspot and some plage 
regions. The second (Map~B)\cite{L96} is a quiet Sun map and is
used here to explore 
the feasibility of the sub-aperture method with quiet Sun signals.
Fig~\ref{fig:maps} shows the observations and the three
slit positions that have been used as solar polarization sources (sunspot,
plage and network).

Let us discuss the numerical experiments in some detail. We start by taking
some spectra from one of the slit position marked 
in Fig~\ref{fig:maps}. These will be our solar input Stokes vectors
from the first observation [$\vec S_{solar,1}^{in} (\lambda,y)$, where $y$
denotes spatial position along the slit and $\lambda$ represents the
spectral variable]. An observation through the telescope sub-aperture is
simulated as:
\begin{equation}
\vec S_{solar,1}^{out} (\lambda,y) = {\bf (xt)} \vec S_{solar,1}^{in}
(\lambda,y) \, .
\end{equation}
The variation of ${\bf (xt)}$ with wavelength is slow and can be neglected
over the spectral range recorded by the detector (typically a few \AA
). 

A second set of Stokes spectra $\vec S_{solar,2}^{in} (\lambda,y)$ is obtained
from the same map, but at a slightly different slit position. Let us denote
the offset between the two observations by $\eta_x$. These vectors are
observed using the full aperture:
\begin{equation}
\vec S_{solar,2}^{out} (\lambda,y) = {\bf (XT)} \vec S_{solar,2}^{in}
(\lambda,y) \, .
\end{equation}

Random noise is added to both $\vec S_{solar,1}^{out}$ and $\vec
S_{solar,2}^{out}$. The noise has a normal distribution and has a larger
amplitude for the case of $\vec S_{solar,1}^{out}$ as a result of the 
smaller observing aperture ($d_0$). I have simulated an exposure time of
$\sim$4~seconds 
in both cases, which results in a noise amplitude of $5\times 10^{-4} / d_0$
(with $d_0$ given in meters) for $\vec S_{solar,1}^{out}$ and $1.25\times
10^{-4}$ for $\vec S_{solar,2}^{out}$. Obviously, it would be possible to use
longer exposure times, thus reducing the noise in the observations. 

At this point we have the simulated solar observations and
the telescope matrices used to generate them. Next we need to apply the
procedure described in \S\ref{sec:proc} to the simulated data and compare the
retrieved matrix ${\bf (XT)'}$ to the actual ${\bf (XT)}$ generated earlier.

In a real situation, one starts by using the calibration
optics to obtain the Mueller matrix (or at least an approximation) of the
sub-aperture system ${\bf (xt)}$. The matrix ${\bf
  (xt)'}$ resulting from the measurements should be a good
approximation to the actual ${\bf (xt)}$. For instance, the ASP calibration
yields an accuracy\cite{SLMP+97} better than $5\times 10^{-4}$. 

The tests presented in this section consider that:
\begin{equation}
{\bf (xt)'} = {\bf (xt)} + \eta_t \sigma \, ,
\end{equation}
where $\eta_t$ is a parameter that accounts for the accuracy of the
sub-aperture calibration (e.g., for ASP $\eta_t < 5\times 10^{-4}$, as stated
above) and
$\sigma$ is again a matrix of random elements. Several experiments with
different values of $\eta_t$ have been carried out in order to explore the
propagation of errors from the calibration of the sub-aperture to the full
aperture system (see below).

Finally, one needs to perform steps~8 and~9 of the procedure detailed in
\S\ref{sec:proc}. A least-squares algorithm is employed to determine the
${\bf (XT)}$ matrix by solving the following system:
\begin{equation}
\label{solution}
\vec S_{solar,2}^{out} = {\bf (XT)} \vec S_{solar,2}^{in} \, .
\end{equation}
Instead of using all the possible Stokes vectors observed at every
$(\lambda, y)$ point along the slit, it is numerically convenient to select a
suitable less redundant 
subset. The strength of the polarization signal of the $\vec
{S'}_{solar,1}^{in}$ is used as a selection criterion. A total of 280 vectors,
consisting of the 40 points with the largest positive values of Q, -Q, U, -U,
V, -V plus the 40 points with lowest total polarization, are
selected. Although the number of vectors has been chosen rather
arbitrarily, the procedure is not very sensitive to it. This value was
adopted after verifying that, in selected simulations, doubling it resulted
in no accuracy improvement.

The least-squares fitting results in a ${\bf (XT)'}$ matrix that can be
straightforwardly compared to ${\bf (XT)}$. In this
manner it is possible to study the propagation of errors and the overall
accuracy of the sub-aperture method. The entire process has
been repeated 100 times for each one of the simulations, with different
realizations of the random matrices that appear in Eqs~(\ref{random}). The
results presented below are the mean-square errors from the montecarlo
simulations.

\subsection{Results}
\label{sec:results}

Let us start the discussion in this section with a simple
reference case. Consider for instance a 
diameter of $d_0=0.30$~m for the calibration optics and a calibration
accuracy $\eta_t = 5\times 10^{-4}$. The solar source is a sunspot (slit
position 1 in Fig~\ref{fig:maps}). Let us assume for the moment that the
telescope pointing has not changed between the two observations ($\eta_x =
0$). The propagation of errors and noise through the calibration procedure
(as explained above) leads to the following error matrix ${\bf E} = 
{\bf (XT)'} - {\bf (XT)}$:
\begin{eqnarray}
\label{referenceerr}
{\bf E} = \left ( \begin{array}{cccc} 
  6.9\times 10^{-4} & 1.3\times 10^{-3} & 2.4\times 10^{-2} & 7.1\times 10^{-4} \\
  6.0\times 10^{-4} & 1.4\times 10^{-3} & 2.3\times 10^{-2} & 7.2\times 10^{-4} \\
  6.2\times 10^{-4} & 1.2\times 10^{-3} & 2.5\times 10^{-2} & 7.3\times 10^{-4} \\
  6.8\times 10^{-4} & 1.3\times 10^{-3} & 2.3\times 10^{-2} & 7.1\times 10^{-4} \\
	    \end{array}  \right ) \, .
\end{eqnarray}

Notice that the second and third columns, which correspond to cross-talk from
Q and U, are larger than the other two. This is due to the fact
that the linear polarization signals are considerably weaker than I and V,
even in the penumbra where the linear polarization is strongest. Fortunately,
the ATST polarimetric accuracy requirement 
(the amount of cross-talk must be smaller than $5\times 10^{-4}$)
does not necessarily imply that the matrix elements of ${\bf E}$ have to be
all smaller 
than $5\times 10^{-4}$. Typically, V is $\sim 10^{-1}I_c$ (where $I_c$ is the
continuum intensity) and Q,U are $\sim
10^{-2}I_c$. This means that, in order to meet this requirement, the elements
in the first column of ${\bf E}$ must be smaller than $5\times 10^{-4}$,
those in the second and third smaller than $5\times 10^{-2}$, and those
in the fourth smaller than $5\times 10^{-3}$ (see reference \cite{E04} for a
more detailed discussion). These values are used as derived requirements for
the ATST polarimetric accuracy in the discussion below.

\begin{figure*}
\includegraphics[width=1.0\textwidth]{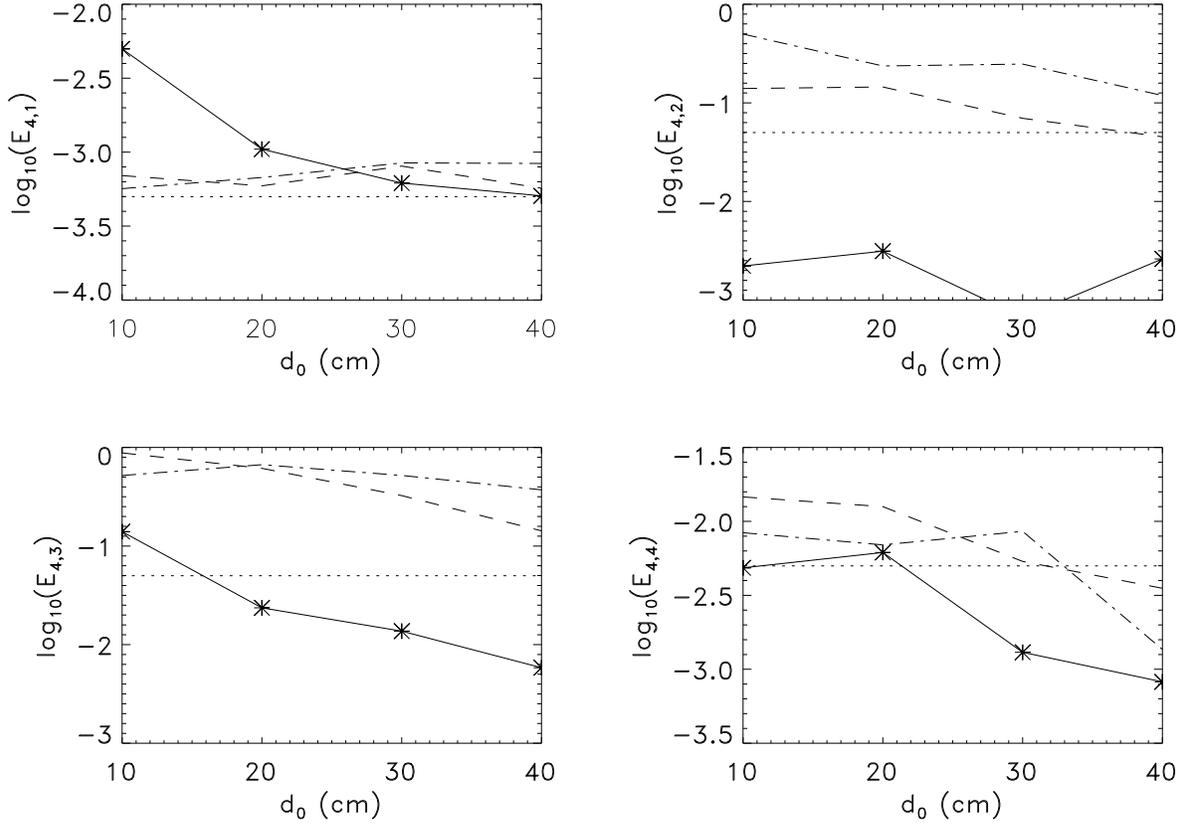}
\caption{Elements of the error matrix ${\bf E}$ as a function of $d_0$ from
  numerous montecarlo hound-and-hares tests. The various lines represent
  different sources of polarization. Solid: Sunspot (slit position~1 in
  Fig~1). Dashed: Plage (slit position~2 in Fig~1). Dash-dots: Network
  element in the quiet Sun (slit position~3 in Fig~1). The horizontal dotted
  line represents the ATST polarimetric accuracy requirements.
\label{fig:tests1}
}
\end{figure*}

Fig~\ref{fig:tests1} shows how the calibration error varies with the
sub-aperture diameter $d_0$ for $\eta_t = 5\times 10^{-4}$ and $\eta_x =
0$. The horizontal dotted line marks the ATST accuracy requirements for each
column of ${\bf (XT)}$. Cross-talk terms from I are almost at the required
level for $d_0 > 20$~cm (upper left panel). A slightly lower
$\eta_t$ would be needed to fully meet the ATST requirement (see below). The
situation is 
better for the other Stokes parameters, but only when using a sunspot as the
polarization source. Plage and quiet Sun fields can be used to determine the
first and fourth columns of ${\bf (XT)}$ at ATST levels. However, the second
and third columns (cross-talk from Q and U) exhibit very large
inaccuracies. The 
reason for this is that the observations used here have been taken at disk
center and show very little linear polarization. In order to fully determine
the Q and U terms of the telescope matrix from plage or quiet Sun regions,
one would also need near-limb observations which exhibit linear polarization
signals.

\begin{figure}
\includegraphics[width=1.0\textwidth]{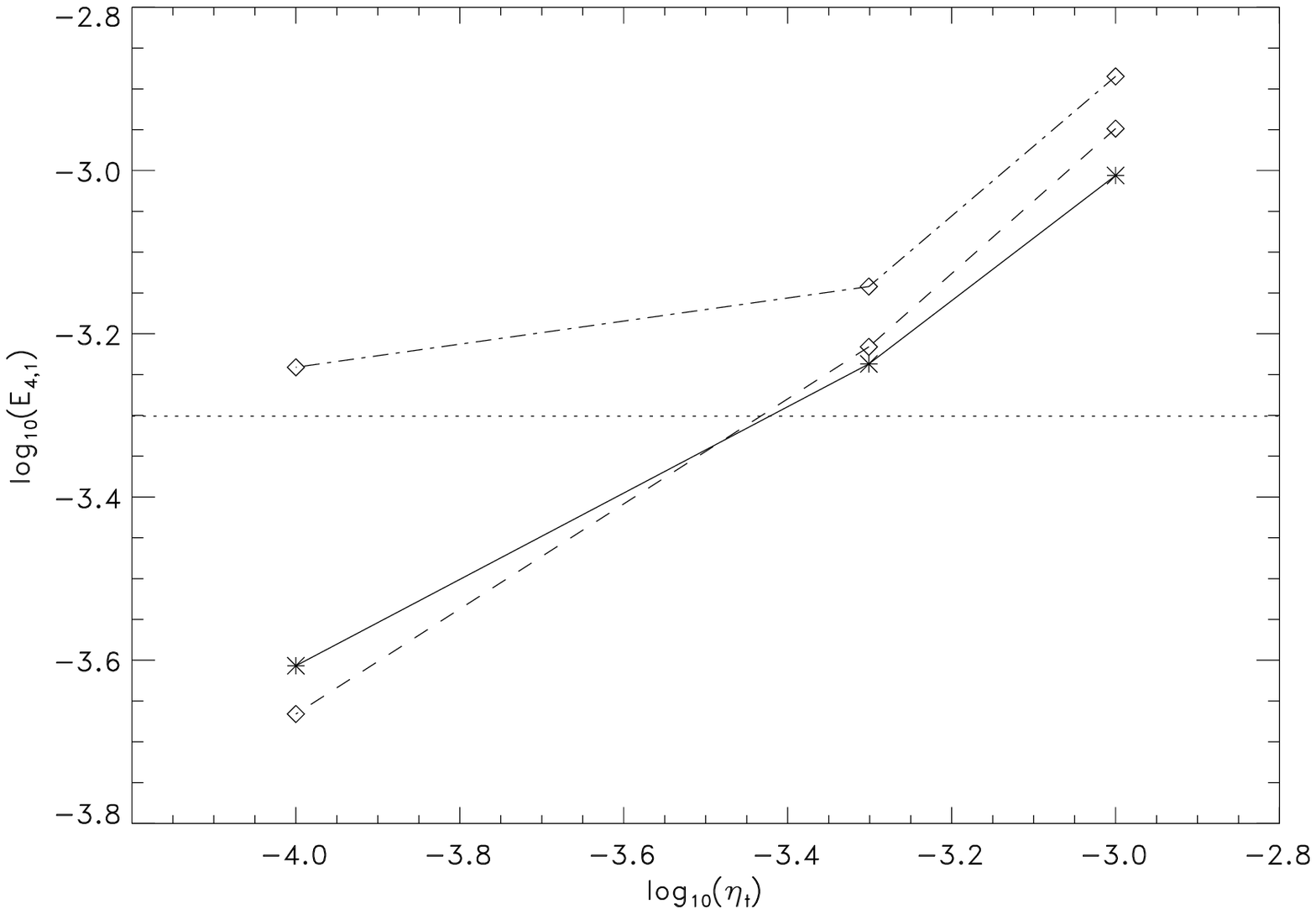}
\caption{Cross-talk from I to V as a function of the sub-aperture calibration
  accuracy ($\eta_t=0$). Solid: No pointing error ($\eta_x$). Dashed:
  $\eta_x$=0.75 arc-seconds. Dash-dots: $\eta_x$=1.5 arc-seconds. The
  horizontal dotted line represents the ATST polarimetric accuracy
  requirement. 
\label{fig:tests2}
}
\end{figure}

It is possible to achieve the required accuracy by calibrating the
sub-aperture to the level of $\eta_t \simeq 3\times
10^{-3}$, as shown in Fig~\ref{fig:tests2}. Notice in that figure that the
telescope calibration accuracy is strongly related to that of the
sub-aperture. Also in Fig~\ref{fig:tests2} we can see the uncertainties
introduced by pointing errors between the observation of $\vec S_{solar,1}$
and $\vec S_{solar,2}$. The method is not compromised by errors as large
as 0.75~arc-seconds and is therefore robust enough to be employed even in the
absence of adaptive optics.

\section{Advantages and Disadvantages}
\label{sec:advantages}

The sub-aperture method has the following advantages over the BEM and MEF:
\begin{itemize}
\item Unlike MEF, no {\it a priori} assumptions are made on the observed
  targets (physical model, profile shape or symmetry, etc).
\item Simpler design modifications than BEM.
\item Unlike MEF, can be employed at any observable wavelength.
\item Unlike BEM, uses the exact same optical setup as the actual observations.
\item Unlike BEM, the full optical train is calibrated, from the primary
  mirror to the detectors. 
\item Unlike MEF, can be employed in the absence of sunspots (e.g., looking
  at plage regions or any other suitable source of polarization vectors).
\item Unlike MEF, can be applied to non-solar telescopes.
\end{itemize}

There are also some drawbacks that should be mentioned:
\begin{itemize}
\item Unlike MEF, requires design modifications.
\item Unlike BEM, requires an astronomical source of polarization.
\end{itemize}

\section{Conclusions}

The sub-aperture method proposed in this work provides a suitable means of
calibrating a large telescope. This paper explores its possibilities and
limitations by means of hound-and-hares tests. Potential advantages and
disadvantages with respect to two other alternatives (the BEM and MEF) are
discussed. 

The simulations indicate that the sub-aperture method is robust enough to
meet the stringent ATST polarimetric requirements in practical situations.
The amount of time required for the calibration procedure is probably
inferior to $\sim$20~minutes. Most of the time is spent observing the
various configurations of the calibration optics, which is the same
procedure employed at the DST (Sacramento Peak) or the VTT (Tenerife) for the
polarimeter calibration. This is done currently on a routine basis about once
per day during the observing runs. 

\section{Acknowledgments}
This work utilizes data from the Advanced Technology Solar
Telescope (ATST) project, managed by the National Solar Observatory,
which is operated by AURA, Inc. under a cooperative agreement with the
National Science Foundation.


\end{document}